\documentclass[journal]{IEEEtran}
\usepackage{natbib}
\usepackage{ifpdf}
\usepackage{cite}
\usepackage{graphicx}
\usepackage{amsmath}
\usepackage{algorithmic}
\usepackage{array}
\usepackage{url}
\usepackage{booktabs}
\usepackage{amsmath}
\usepackage{multirow}
\usepackage{txfonts}
\usepackage{multicol}
\usepackage{booktabs}
\usepackage{mathdots}
\usepackage{amssymb,amsfonts}
\usepackage{textcomp}
\usepackage{xcolor}
\usepackage{helvet}
\begin{document}
	
	\title{Early Prediction of Epilepsy Seizures VLSI BCI System}
	
	\author{Zaghloul Saad Zaghloul$^{\dagger}$ \qquad Magdy Bayoumi$^{\star}$\\
		\textit{$^{\dagger}$Center for Advanced Computer Studies} \\
		\textit{$^{\star}$Electrical and Computer Engineering Department} \\
		\textit{$^{\dagger}$$^{\star}$University of Louisiana at Lafayette}\\
		$^{\dagger}$$^{\star}$Lafayette, Louisiana, USA\\
	}


	\maketitle

	\begin{abstract}
		Controlling the surrounding world and predicting future events has always seemed like a dream, but that could become a reality using a Brain-Computer/Machine Interface (BCI/BMI). Epilepsy is a group of neurological diseases characterized by epileptic seizures. It affects millions of people worldwide, with 80\% of cases occurring in developing countries. This can result in accidents and sudden, unexpected death. Seizures can happen undetectably in newborns, comatose, or motor-impaired patients, especially due to the fact that many medical personnel is not qualified for EEG signal analysis. Therefore, a portable automated detection and monitoring solution is in high demand. Thus, in this study, a system of a wireless wearable adaptive for early prediction of epilepsy seizures is proposed, works via minimally invasive wireless technology paired with an external control device (e.g., a doctors’ smartphone), with a higher than standard accuracy 71\% and prediction time (14.56 sec). This novel architecture has not only opened new opportunities for daily usable BCI implementations, but they can also save a life by helping to prevent a seizure’s fatal consequences.
	\end{abstract}
	

	\IEEEpeerreviewmaketitle
	\section{Introduction}
	Epilepsy is a group of neurological diseases characterized by epileptic seizures that affect more than 10\% of the human population worldwide; nearly 80\% of cases occur in the developing world, and resulted in 116,000 Sudden Unexpected Death in Epilepsy (SUDEP) in the last two years,~\citet{george2014design}. Epilepsy becomes more common as people age. In developed countries, infants account for most of the seizures. About 5-10\% of people over 80 years old have had a seizure. Sufferers have an increased chance of experiencing a second seizure, and usually, epilepsy cannot be cured by~\citet{hitiris2007mortality}. Epilepsy can be a primary cause of sudden death or cause different accidents, especially motor-vehicle accidents,~\citet{marx2013rosen}. Seizures can also happen to newborns, comatose, or significant motor-impaired patients, especially in Intensive Care Units (ICU). The situation can be even more severe when medical personnel is not qualified for EEG signal analysis, which is a very common case. Using online unsupervised BCI, detection and monitoring the solution can at least help in detecting the seizure symptoms early to avoid fatal consequences,~\citet{eadie2012shortcomings} and~\citet{longo2012369}.

Despite the advances in pharmacological treatments, approximately 1 in 3 patients continues to experience frequent seizures. Usually epilepsy cannot be cured; however, medication can control seizures effectively, although in more than 30\% is not effective in a generalized seizure and 50\% of people with partial seizures. Dan, 2012 showed that in the developing world, 75\% of people are not appropriately treated, especially in Africa, where 90\% of the patients do not get treatment at all; the global distribution of seizure cases is shown in Fig~\ref{figure_3}.

The current seizure detection solutions are based on manual inspection, and the demand for automated detection is very high, not to mention the need for prediction. The current state-of-the-art architectures either require a perfect environment to operate, as they are prone to noise, have limitations with predictions due to the static techniques used, or require a powerful computational machines as well as a lot of wiring between the patient and the monitoring station, and in most of the cases, they do not provide a closed loop prediction and detection systems nor easy doctor and ICU personal access to decisions making mechanisms, which in many cases can cost the patient’s life.

\begin{figure*}%*[t]
	\centering
	\includegraphics[width=13cm,height=6cm]{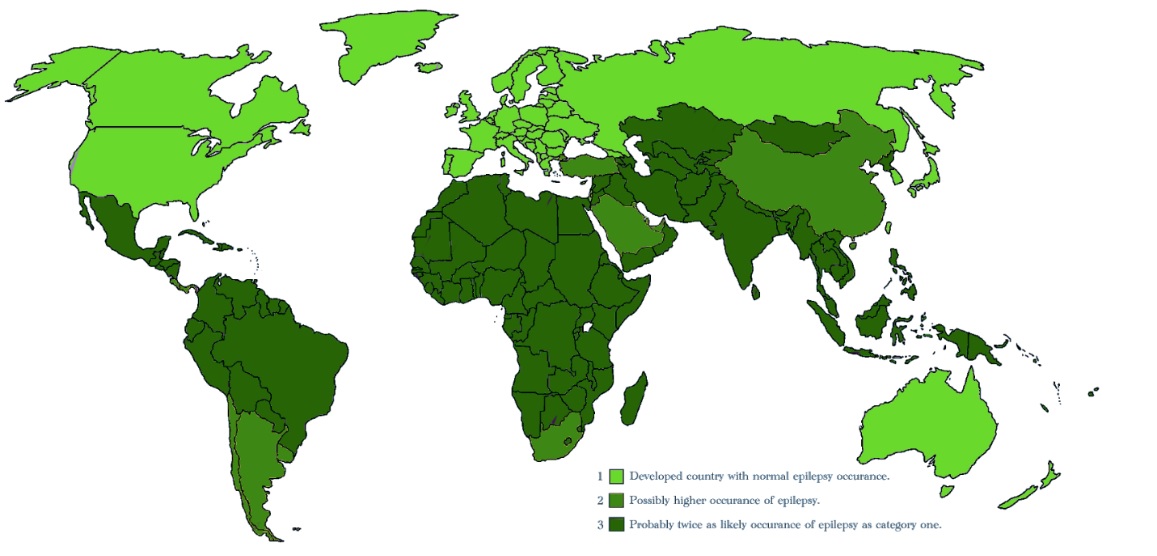}
	\caption{The global epilepsy rate research, the University of Oxford in Lancet.}
	\label{figure_1}
\end{figure*}

In~\citet{labiner2010essential}, a study shows that in most cases the ICU personnel is not qualified for EEG analysis and the majority of patients in the ICU are comatose or have a significant motor impairment. The main problem with the current detection systems is that they are only for measuring continuous EEG monitoring devices that are being reviewed by neuroscientists once or twice a day. These devices usually require a lot of wiring, and they do not provide an early prediction or detection feature. Instead, they depend solely on the visual inspection of the printed EEG waveform activities. However, the limitations of the detection time are a major challenge. Wireless scalp EEG based early prediction, warning and detection of epilepsy seizure systems were proposed and mentioned by~\citet{rogowski1981prediction} and~\citet{rajdev2010real}. According to and WHO 2016, a detection system can be:
\begin{itemize}
	\item Tolerant to noise BCI artifacts and equipment surrounding noise for the ICU room device and the adjacent EMF cross talk and coupling,
	\item Easy to be used; deployed on different patients from different age groups, and different injury of sickness level, as well as connected to warning and alerting device, whether it is a doctor’s smart device machine, ICU monitoring device, or a cloud-based analysis unit.
	\item Provides adequate accuracy (more than 60\%) with respect to the agreed seizure detection standard,~\citet{stacey2008technology}.
	\item Capable of predicting even earlier in the stages of epilepsy seizures, than the current state-of-art (6.4 sec),
	\item Minimizing false-positive alarms (VLSI architecture) to be implemented on a chip and a part of the wearable device.
\end{itemize}

\begin{figure}%*[t]
	\centering
	\includegraphics[width=8.5cm,height=5cm]{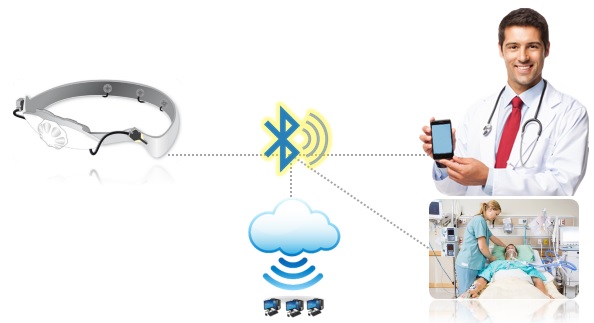}
	\caption{The proposed system conceptual diagram.}
	\label{figure_2}
\end{figure}

Many proposed seizure detection systems require an “idea signal acquisition,” which makes them perform poorly under the presence of additional noises and BCI artifacts,~\citet{nicolas2012brain}. In \citet{castro2002artificial} defined Artificial Immune Systems (AIS) as adaptive problem-solving systems, inspired by theoretical biology and immunology functions, Artificial Immune Systems (AIS) are an adaptive clustering and classifier system that is used to detect the abnormality within problem sample space, based on an ideal data set, and the working space of regenerates and mutated samples to compare. These give the AIS a promisingly accurate prediction of behavior, especially for rouge types of input patterns. Therefore, in this study, we propose an early predictor of epileptic seizures that is based on disposable wearable non-invasive sensors placed on a headband that communicates with a smartphone or any ICU monitoring device via a Bluetooth connection technology, shown in Fig~\ref{figure_2}.

The proposed system uses a minimally invasive wearable EEG band, with a signal processing chip attached to it, which is placed behind the patient’s ear and connects via Bluetooth technology to an external monitoring device, or to ICU personnel or doctor’s smart handheld device. The results, which include sensitivity and duration parameters configuration are also given to the doctors to set for custom configuration via the prediction smartphone software. It allows the uploading of the raw data and the seizure ictal events recording to the cloud for further research, evidence recording, or tuning the initial population of signature for the detection algorithm. The proposed system is capable of early prediction by an average of 14.56 sec, especially for the long-lasting seizures (called: status epilepticus by~\citet{rogowski1981prediction}, with accuracy more than 71\% higher than the agreed standards (>60\%), giving enough time (>10sec) for warning and treatment,~\citet{haas2007strategies}, and~\citet{jerger2001early}. This means our proposed systems can bring back mobility of a limb (artificial or biological) to a handicapped patient or can save lives through early prediction of a seizure.

%%%%%%%%%%%%%%%%%%%%%%%%%%%%%%%%%%%%%%%%%%%%%%%%%%%%%%%%%%%%%%%%%%%%

\section{Background}
\subsection{Epilepsy seizure types}
\begin{figure*}%*[t]
	\centering
	\includegraphics[width=14cm,height=10cm]{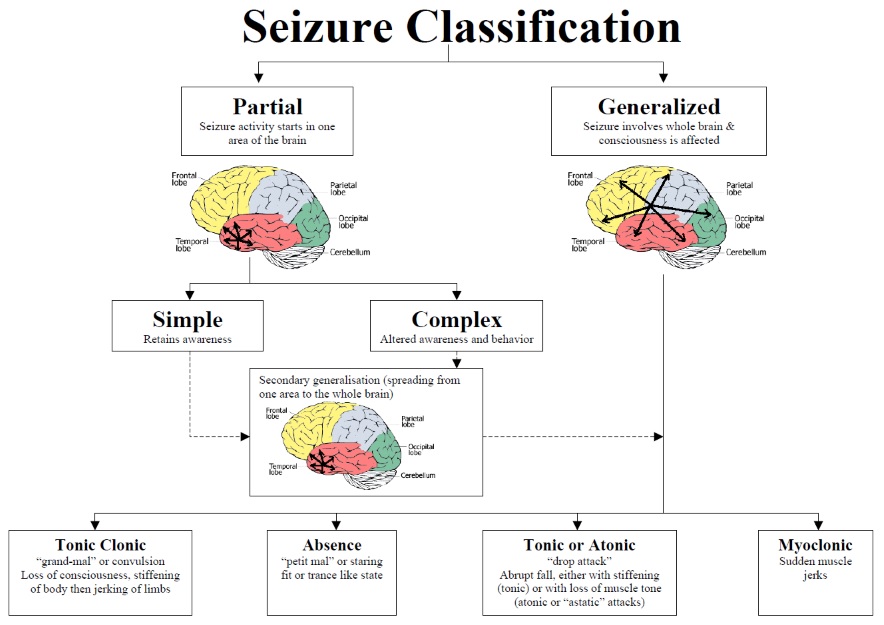}
	\caption{Epileptic seizure classification~\citet{fisher2014ilae}.}
	\label{figure_3}
\end{figure*}

According to ~\citet{roger2000clinical} all the seizures could be divided into two main categories: Partial Seizure and Global Seizure, where it can be subdivided into interlaced subcategories based on the location of the seizure, the patient’s visual symptoms, and the awareness state, Fisher et al., 2014; the classification can be categorized as shown in Fig~\ref{figure_3}. 

\begin{figure}%*[t]
	\centering
	\includegraphics[width=8cm,height=5cm]{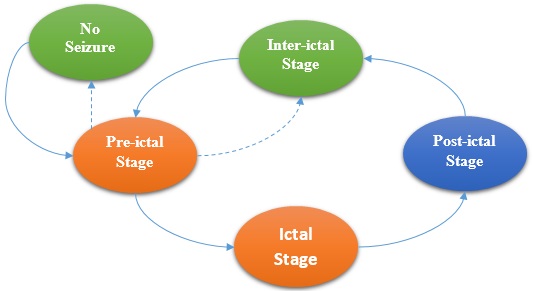}
	\caption{Epilepsy seizure model state diagram.}
	\label{figure_4}
\end{figure}

\citet{castro2002artificial} proposed regarding seizure detection and for seizure prediction. Additionally,~\citet{cecotti2011convolutional} divided seizures into several states. The seizure state diagram is shown in Fig~\ref{figure_4}:
\begin{itemize}
	\item \textbf{Ictal}: A physiologic state during a seizure recording. Latin ictus, meaning a blow or a stroke.
	\item	\textbf{Pre-ictal}: The state immediately before the actual seizure or sometimes the beginnings of the ictal state.
	\item	\textbf{Post-ictal}: The state shortly after the seizure.
	\item	\textbf{Inter-ictal}: The period between seizures, or convulsions, that is characteristic of an epilepsy disorder. For most people with epilepsy, it inter-ictal occupies more than 99\% of their lifetime.
\end{itemize}

An epileptic seizure is defined as a brief episode of signs or symptoms due to abnormal excessive or synchronous neuronal activity in the brain that can result in uncontrolled movement of the limbs and almost any muscle of the body, and it usually is accompanied by a loss or impairment of awareness. It is a disease of the brain characterized by an enduring predisposition to generate epileptic seizures, which are collectively called epilepsy.~\citet{le2001anticipation} showed that all epileptic seizures can be classified into six main types:
\begin{itemize}
	\item \textbf{Tonic-clonic seizures}: characterized by a noticeable contraction of the limbs followed by their extension, along with arching of the back for 10–30 seconds, during which a cry may be heard due to contraction of the chest muscles.
	\item \textbf{Tonic seizures}: results in a constant contraction of the muscles. The person may even turn blue if breathing is impaired.
	\item \textbf{Clonic seizures}: a form of shaking of the limbs; ranges from simple to severe shaking.
	\item \textbf{Myoclonic seizures}: usually cause spasms of muscles in either a few areas or are generalized through the body.
	\item \textbf{Absence seizures}: results in only a slight turn of the head or eye blinking where the person often does not fall over, but its post-ictal can last for hours after the seizure ends.
	\item \textbf{Atonic seizures}: loss of muscle activity for greater than one second; this typically occurs on both sides of the body.
\end{itemize}

\subsection{Seizure and EEG}
\citet{krumholz2007practice} defined a seizure as changes in behavior that occur after an episode of abnormal electrical activity in the brain. It could be also defined as convulsions that occur when a person's body shakes rapidly and uncontrollably. During convulsions, the person's muscles contract and relax repeatedly. There are many different types of seizures, even those that have symptoms without shaking.~\citet{luders2000atlas} provided an example of different types of seizures and the way that they appear on the EEG recordings, which is basically analyzed via visual inspection by a neuroscientist. First of all, there is a snapshot of the EEG monitoring of a medical test done at the Department of Neurology, University of Florida and the Atlas for Electroencephalography.

\begin{figure*}%*[t]
	\centering
	\includegraphics[width=\textwidth,height=4cm]{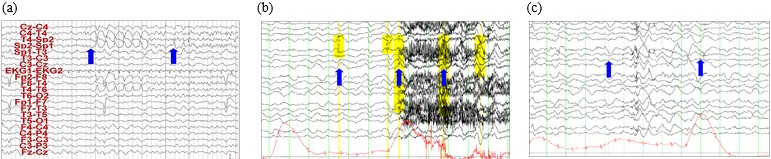}
	\caption{(a) Right temporal seizure maximal phase reversal in the right temporal lobe; (b) Nocturnal Frontal Lobe Seizure hidden activity; (c) Nocturnal Frontal Lobe Seizure,~\citet{luders2000atlas}.}
	\label{figure_5}
\end{figure*}

The partial seizure results are basically no loss of awareness, it can take place at the focal motor movement. It can happen at the temporal lobe (e.g. smell, epigastric sensation, Deja vu at the hippocampus, fear or anxiety at the amygdala), at the Parietal lobe (e.g. sensory), or at the Occipital lobe (e.g. visual). Additionally, a supplementary Motor Seizure can be combined with dystonic posturing of the upper extremities (fencing) or lower extremities (Bicycling) and it usually lasts for relatively short duration 10-30sec, shown in Fig~\ref{figure_5}(a).

Another type of common case seizure is the Autosomal Dominant Nocturnal Frontal Lope Epilepsy Seizure (ADNFLE). It is usually a brief (<2 minutes) thrashing behavior similar to arising from sleep position. An interesting point is that during that type of seizure the patients may be observed like normal; this is the main reason why for a long period of time it was diagnosed as a simple movement of a sleep disorder. An example of the recorded EEG is shown in Fig~\ref{figure_5}(b) and Fig~\ref{figure_5}(c).

\begin{figure*}%*[t]
	\centering
	\includegraphics[width=\textwidth,height=5cm]{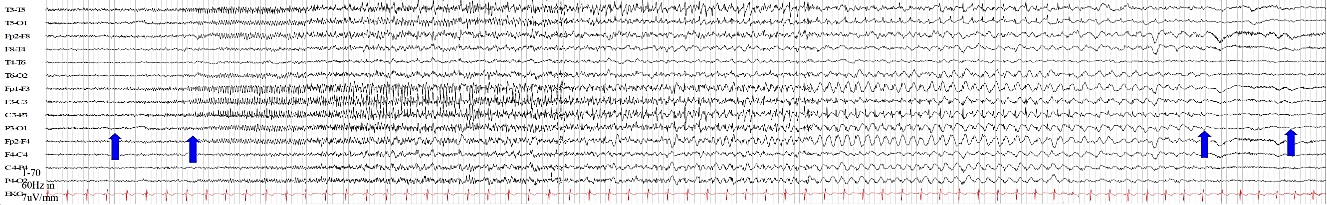}
	\caption{The full period of a focal generalized seizure.}
	\label{figure_6}
\end{figure*}

The second major class of epilepsy seizure is Generalized Seizures. It usually begins focally, with or without focal neurological symptoms, then it spreads everywhere in a very short time, which makes detecting its origin a very hard problem, Many studies even consider the origin of the Generalized Seizure to be unknown in the majority of the cases. The Grand-mal basic characteristic is a variable symmetry, intensity, and a long tonic duration (1-3 minutes) and/or clonic phases. The generalized seizures usually result in a strong and long lasting post-ictal symptom of confusion, somnolence, with or without a transient focal deficit. An example of the whole generalized seizure from beginning to end is illustrated in Fig~\ref{figure_6}.

\begin{figure*}%*[t]
	\centering
	\includegraphics[width=\textwidth,height=5cm]{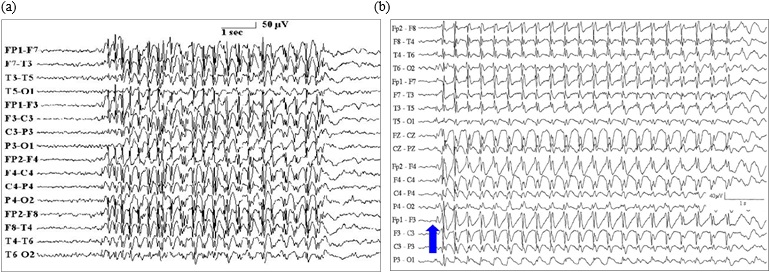}
	\caption{(a) Atypical Absence Seizure EEG Recording; (b) a childhood absence epilepsy shows the synchronization between the channels.}
	\label{figure_7}
\end{figure*}

One of the most widely diagnosed seizures that happen to children and newborns (especially between 4 and 7 years old) is called Absence Seizure. It is usually occupied with a brief impairment of awareness and usually lasts for 3-20 seconds. The main characteristic of the Absence seizure is that it has a sudden onset and sudden resolution. Also, Absence Seizures are often provoked by hyperventilation, this type of seizure usually resolves when the patients enter adult age. Absence Seizures can be easily detected on the 3Hz spike-wave discharges, as shown in Fig~\ref{figure_7}(a) and Fig~\ref{figure_7}(b).

\begin{figure*}%*[t]
	\centering
	\includegraphics[width=\textwidth,height=5cm]{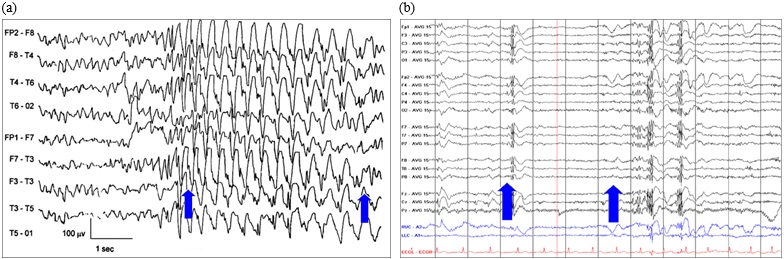}
	\caption{(a) Juvenile Absence Seizures, the arrow are for the beginning and the resume points; (b) Myoclonic Seizures EEG recordings.}
	\label{figure_8}
\end{figure*}

Juvenile Absence Seizures are a special case of Absence Seizure; they are characterized by brief starting spells with variably reduced responsiveness that lasts on average 5-30 seconds, and usually occurs in children (7-8 years of age). It does not fade away as the previously describe a seizure. However, it is very less frequent than the onset of Absent Seizures, shown in Fig~\ref{figure_8}(a).

Myoclonic Seizures are brief, shock-like jerks of a muscle or group of muscles. This type of seizure is usually characterized by bilaterally synchronous impairment of consciousness and short activity bursts (<1 second). The EEG recordings are demoralized over the 4-6 Hz frequency range with poly-spike-wave discharges, as shown in Fig~\ref{figure_8}(b).

\begin{figure*}%*[t]
	\centering
	\includegraphics[width=\textwidth,height=10cm]{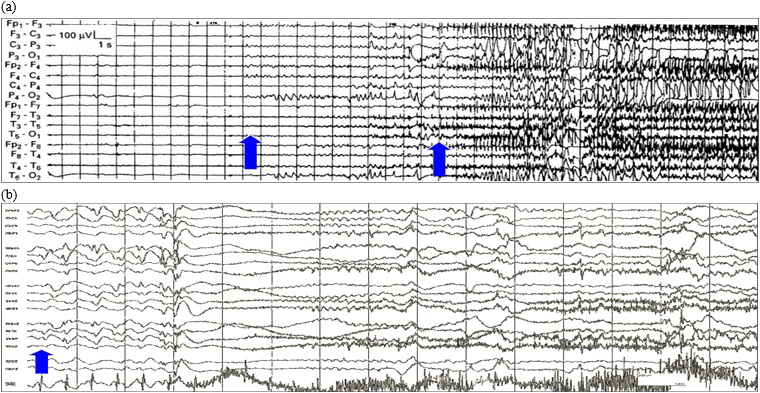}
	\caption{(a) Tonic seizures: basically results in a constant contraction of the muscles, and even the person may turn blue if breathing is impaired; (b) M•	Tonic-Clonic seizures: can be characterized by a noticeable contraction of the limbs followed by their extension, along with arching of the back for 10–30 seconds, during which a cry may be heard due to contraction of the chest muscles.}
	\label{figure_9}
\end{figure*}

Tonic Seizures are symmetric, tonic muscle contraction seizures that usually result in tonic flexing of the waist and the neck. It usually lasts for 2-20 seconds. The recorded EEG shows a sudden attenuation with generalized, low-voltage fast activity, or a generalized poly-spike-wave. An example of a Tonic Seizure is shown in Fig~\ref{figure_9}(a), and a special case of a Tonic-Clonic Seizure is shown in Fig~\ref{figure_9}(b).

The next section provides a review of the existing works in BCI and automated seizure detection, prediction tool, approaches, and technology.

%%%%%%%%%%%%%%%%%%%%%%%%%%%%%%%%%%%%%%%%%%%%%%%%%%%%%%%%%%%%%%%%%
\section{Related work}

\subsection{Statistical Detection}

\begin{figure}%*[t]
	\centering
	\includegraphics[width=8cm,height=6cm]{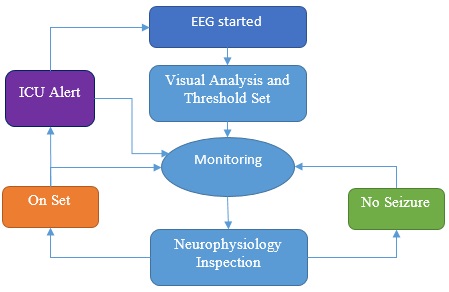}
	\caption{The synchronization likelihood test system flow chart.}
	\label{figure_10}
\end{figure}

Unlike in~\citet{maynard1969device} the Amplitude Integrated EEG (aEEG) is widely used to detect neonatal seizures, despite the fact that the accuracy of seizure recognition can be moderate, especially in brief, low amplitude, focally oriented seizures,~\citet{young2009seizure}. Neonatologists analyze the aEEG signals at the patient's bedside. In our opinion, clinical neurophysiologists should at least be involved in this interpretation, as they are specially trained in EEG reviewing. Where they compared a four-channel EEG monitoring device with 16 channel EEG recordings and found a sensitivity of 68\% and specifity of 98\% with a visual interpretation of the signals; the statistically based algorithm is shown in Fig~\ref{figure_10}.

In~\citet{ponten2010feasibility} has experience with another quantitative analysis approach for EEG, namely the synchronization likelihood (SL), where a nonlinear measure of statistical interdependencies between time series, which has shown to be a promising measure for detecting seizures in neonatal EEGs and frontal lobe epilepsy. They used an SL based technique to design an online automatic detection method for real-life EEG seizure monitoring. The experiment used the EEG data from 50 patients over a 15-month period, measuring how five were suffering from seizures. Using the 16 channel EEG, the architecture was able to detect most of the seizure with 68\% sensitivity and specificity of 98\%.

Despite the accurate results and faster performance, this architecture suffers from several drawbacks that limit its implementation as a wearable device. The first disadvantage is that it requires a near ideal operation scenario as this design is prone to noise and BCI artifacts. The second disadvantage, which is mentioned by the authors, is that the frame structure scanned could produce a high rate of false positives especially in the case of a scalp EEG input electrode.

\subsection{SVM Base Seizure Detection Techniques}

\begin{figure*}%*[t]
	\centering
	\includegraphics[width=\textwidth,height=5cm]{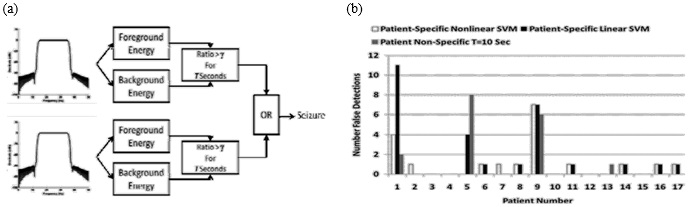}
	\caption{(a) Low-power SVM bases detection algorithm; (b) the system detection result graph,~\citet{roger2000clinical}.}
	\label{figure_11}
\end{figure*}

There are several works that focus on Epileptic Seizure Detection. ~\citet{roger2000clinical} addresses the computational and implementation challenges associated with detecting seizure onset with an implantable device. The study specifically shows how a Two-Class Support Vector Machine (SVM) can be used to synthesize patient-specific detectors that outperform a patient non-specific detector. The study also discusses other methods that enable efficient implementation of the discriminator functions produced by the SVM algorithm. The system circuit diagram and simulated results are shown in Fig~\ref{figure_11}.

The researchers claim that: ``the detector extracts, from each channel, features that can be used to infer the presence of seizure activity. Since spectral energy has been shown to be useful in the context of intracranial seizure detection, the chosen features where the energy within the frequency bands 0-16Hz and 15-37Hz''.

Regardless of the low power achievements this design has several drawbacks, including that the use of only static patient related thresholds for seizure detection, requires invasive implants, which are very sensitive to noise and almost fits the lower bound of the acceptable false alarms.

\subsection{Pattern Recognition Based Detection}

\begin{figure}%*[t]
	\centering
	\includegraphics[width=8.5cm,height=6cm]{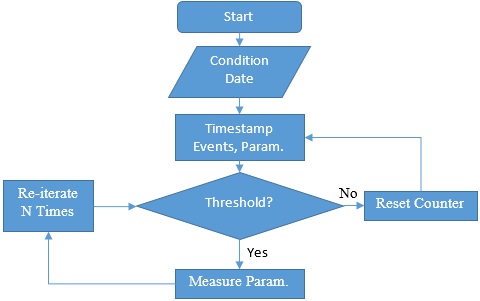}
	\caption{Flow chart of the operation stage of the algorithm.}
	\label{figure_12}
\end{figure}

\citet{ponten2010feasibility} present a novel event-based seizure detection algorithm along with a low-power digital circuit implementation [76]. Using invasive depth-electrode, recordings from rats are used to validate the algorithm and hardware performance. The main detection algorithm is shown in Fig~\ref{figure_12}.

The research also discusses the main problems and cost in translating mathematical models into hardware implementations. It also claims they could reach an average TPR of 95.3\% and FPR of 88.9\%. They implemented their system on a CMOS circuit drawing ~350nW of a power source with 250mV.

\subsection{Adaptive Prediction Algorithm}

\begin{figure}%*[t]
	\centering
	\includegraphics[width=8.5cm,height=5cm]{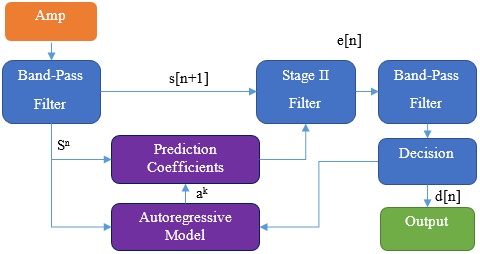}
	\caption{Block diagram of the real-time algorithm.}
	\label{figure_13}
\end{figure}

\citet{rajdev2010real} discuss from a signal processing point of view, that a data analysis and prediction scheme can be considered to be a four-step process of signal enhancement, including adaptive autoregressive modeling and prediction, envelope detection and a binomial decision rule. Their proposed architecture is based on scalp EEG electrode for data acquisition. A block diagram of the real-time adaptive seizure prediction algorithm that is based on Wiener algorithm is implemented in real-time Fig~\ref{figure_13}.

After the amplitude stage, the band-pass infinite impulse-response filter  (10-500Hz) attenuates the low-frequency noise and the higher frequency artifacts that could lead to false positive detection of seizures,~\citet{longo2012369}. They accept an adaptive threshold value as given below:

\begin{equation}
T= \gamma \frac{1}{N} \sum_{n=1}^{N} e^{2}[n]
\end{equation}
\indent
where N is the number of samples in the window length, $\gamma$ is a scaling factor, and $e^2[n]$ is the squared prediction error signal.  This signal is then modeled as an autoregressive (AR) or all-pole process and given as an input to the prediction filter. A best fit AR model is then determined and future value of the signal s[n+1] is estimated. The error filter then computes the error between the observed and the estimated value. The envelope of the error is used by the decision rule to make a binary decision. The architecture was able to predict seizure on average 6.4 sec ahead with an accuracy of 64\%.

As a summary, most of the previous works are based on a computational intense algorithm or not suitable for VLSI implementation due to the area and power limitations,~\citet{cecotti2011convolutional}, and most of them are proposed for seizure detection but do not focus on the early prediction of the pre-ictal stage. Additionally, they do not solve the threshold update problem, the artifact noise attenuation, the detection bounded minima problem, the static pattern storage, not the framing bounders problem, respectively.

\section{Proposed System}

\subsection{Proposed Method Concepts}

The proposed system uses a non-invasive wearable EEG band, with a signal processing chip attached to it, which connects via Bluetooth technology to an external monitoring device, ICU personnel or doctor’s smart handheld device. The sensitivity and duration parameters configuration are also given to the doctors to set for custom configuration via the prediction smartphone software (i.e. Phone App.). It also allows data upload of the raw data and differ ictal event recording to the Cloud for further research, evidence recording, or tuning the initial population of signature for the detection algorithm.

\subsection{Cauchy-Based Filter}

Many proposed seizure detection systems require ideal signal acquisition, which makes them perform poorly under the presence of additional noises and BCI artifacts.~\citet{wang2015cauchy} shows that more than 90\% of the affecting noises of the scalp EEG recording comes from the EOG and EMG artifacts, where the rest comes from non-acceptable electrode conductivity, loose electrode connections, the surrounding electrical resonance, adjacent electric and magnetic fields EMF and wire crosstalk.

\citet{wang2015cauchy} studied different noise reduction and elimination techniques to avoid the Gaussian assumption of the stationary noise, and found the best results of false detection rate (FDR) are 0.15/h and 0.08/h for the overall and best case of the experimental scenario, where they had life EEG data from ten human patients. That was achieved via using a Cauchy-based observation together with the autoregressive (AR) model to represent the state transitions that could be modeled as:
\begin{equation}\label{eqn2}
x_t= \sum_{i=1}^{p} \alpha_i x_{i-1} + \beta + \upsilon_t, \quad \upsilon_t \approx N(0, \sigma_{\upsilon}^{2})
\end{equation}
\indent
where $x_t$ represents the binary state of seizure occurrence, over the state transition of order $p$, at time $t$, where $\alpha_i$ and $\beta$ are distribution coefficient and a constant, respectively. Also $\upsilon_t$ is the Gaussian state noise with an orthogonal mean and variance. Regardless of the debate about the noise sources and their independence or correlation degree, it is seen from~\ref{eqn2} that as soon as noise and artifacts invade the EEG signal, especially closer to the tail, the probability value of the detection will be close to zero, as a property of the Gaussian distribution. This is because the Cauchy distribution~\ref{eqn3} has a higher probability density function (PDF) than the Gaussian case.

\begin{equation}\label{eqn3}
f(x; x_0, \gamma) = \frac{1}{\pi \gamma [1+ (\frac{x-x_0}{\gamma})^2]}
\end{equation}
\indent
where $x_0$ is the location parameter. It is specifying the location of the peak of the distribution, and $\gamma$ is the scale parameter (i.e. called the probable error). Obviously, the maximum value or amplitude of the Cauchy PDF is $\frac{1}{\pi \gamma}$ which is located at the peak. Thus, relying on the Cauchy distribution instead of the Gaussian helps to clearly distinguish between misleading the merged signals. Either it is a seizure event or a system of most known artifacts. We also aware that this filtering could be compromised in the case of overlapping several simultaneous artifacts and/or seizure occurrences; however, from literature these scenarios happen are very rare in the usual practice.

\subsubsection{Artificial Immunity System}
\citet{hassan2013artificial} discussed the history of AIS and showed that at the late 1970s showed considerable interest in biology as a source of inspiration for solving computational problems. Models of the central nervous system have driven artificial neural networks. The Darwin Theory spawned evolutionary simulations in natural selection. According to~\citet{dasgupta1993overview} the definition of the biological immune system (IS) is “an accurate and elaborate defense system that consists of multi-layers of protection where each layer provides different types of defense mechanisms to find, detect, recognize and response for foreign organisms and pathogens”.

\subsubsection{Negative Selection Mechanism}
In~\citet{timmis2007artificial} the Negative selection is defended as a process of selection that takes place in the thymus gland. Its main purpose is to provide tolerance for self-cells. It deals with the immune system's capability to detect unknown antigens while not reacting to self-cells. The maturation of the T-cells is very simple. T-cells are exposed to self-proteins in a binding process. If this binding activates the T-cell, then the T-cell is killed, otherwise, it is allowed into the lymphatic system,~\citet{nguyen2012research}.

\subsubsection{AIS Definition}
\begin{figure}%*[t]
	\centering
	\includegraphics[width=8.5cm,height=10cm]{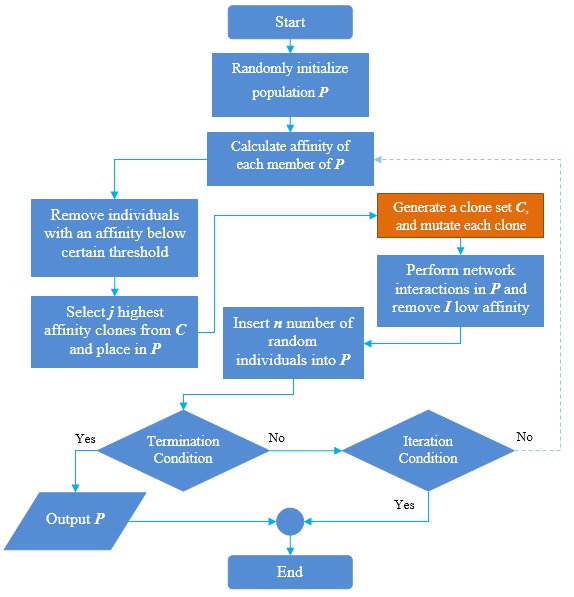}
	\caption{Flow chart of the operation stage of the algorithm.}
	\label{figure_14}
\end{figure}

Artificial Immune System (AIS) is concerned with computational methods inspired by the process and mechanisms of the biological immune system. The scope of artificial immune systems could be thought to be restricted to pattern recognition tasks. The basic negative selection algorithm approach proposed by~\citet{esponda2004formal} as shown in Fig~\ref{figure_14}.

\subsubsection{Neural Signature}
\citet{abu2008vlsi} defined a method of describing a specific neural activity of firing neurons over a specific period of time of the epoch under study that provides a unique characteristic of this activity is defined as a neural signature. The neural signature was an essential step towards the implementation of the parameterized AIS architecture that is the core prediction mechanism of the proposed system. Neural signature is generated via spatial and temporal characteristics of a period of observation not only on the current dominant channel but also taking into consideration the surrounding channels at the same event time $t$ and a slight additional window shift of time $\Delta t$, to cover the neural signal propagation and to solve the moving frame boundary problem.

\subsection{Proposed Model Architecture}
\begin{figure*}%*[t]
	\centering
	\includegraphics[width=11cm,height=9cm]{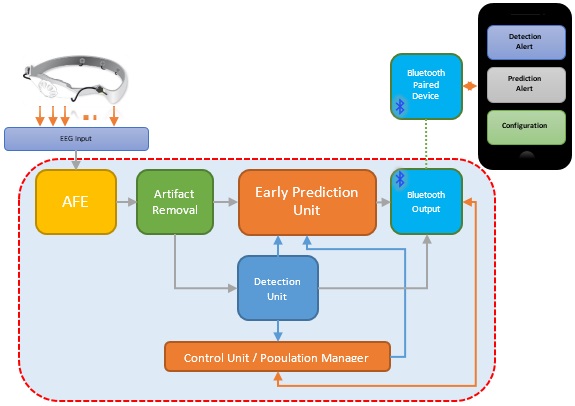}
	\caption{The proposed solution logical block diagram.}
	\label{figure_15}
\end{figure*}

The proposed system uses a non-invasive wearable EEG band with a signal processing chip attached to it and sticks behind the patient’s ear, which connects via a Bluetooth technology to an external monitoring device or an ICU personnel or doctor’s smart handheld device, where the sensitivity and duration parameters configuration is also given to the doctors to set for custom configuration via the prediction smartphone software (i.e. Phone App.) It also allows data upload of the raw data and different ictal events recording to the cloud for further research, evidence recording, or tuning of the initial population of signatures for the detection algorithm. The system solution diagram is shown in Fig~\ref{figure_15}. The gray arrow path represents the input data manipulation path, the blue arrow represents the control lines, the orange arrows represent the parameter configuration lines, and the green dotted line represents the wireless line to the outside control device.

As shown in the previous Fig~\ref{figure_15}, the proposed architecture can logically be divided into three main stages, the Signal Conditioning, the Adaptive Signal Analysis, and the Decision Making stages. The Signal Condition consists of three units: The Analog Front-End (AFE), the Digital Wavelet Transform (DWT), the Signature Generation Unit (SGU), and the Artifact Removal Unit (ARU). Those are followed by the Adaptive Signal Analysis that is consistent with the Seizure Detection Unit (SDU) and the Seizure Prediction Unit (SDU). Finally, it will be followed by the Decision-Making stage that consists of the Decision Controller and the RF Module, which is the gateway between the system on the chip and the control interface on the doctor’s smart handheld device. We will discuss each stage in more details.

\subsubsection{Signal Conditioning Stage}

\begin{figure}%*[t]
	\centering
	\includegraphics[width=8.5cm,height=5.5cm]{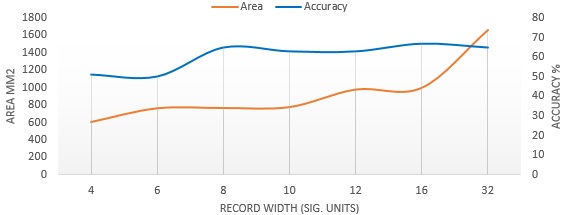}
	\caption{The empirical simulations between record width W, accuracy and estimate area.}
	\label{figure_16}
\end{figure}

This stage is the first part of the online system where the input is the raw EEG from four input channels which are amplified 100 times, then processed via a bandpass filter through a 30-100 Hz band to cut off the high frequency. It is sampled over 250/500Hz inside the AFE unit with a resolution of 16-bit analog to digital converter (ADC) then it is fed to the Cauchy based ARU. We used the unit described in~\citet{wang2015cauchy} as our main artifacts removal block in the adaptive early prediction unit. Then, the signal is sent to the DWT unit to smooth the signal and reduce the sample space dimensions. After W cycles (where W represents the width of signature window) are selected empirically to optimize the chip area and the signature detection of the signal’s amplitude envelope, it is set to 8 Signature Units (Sig.) after the analysis of the simulation results and an adequate value for the window length is shown, such as in Fig~\ref{figure_16}.

\begin{figure}%*[t]
	\centering
	\includegraphics[width=8.5cm,height=6cm]{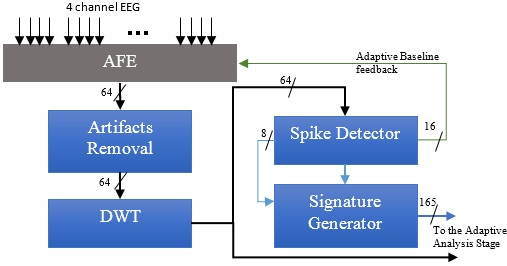}
	\caption{The signal conditioning stage block diagram, with an input of raw EEG data and output of the EEG DWT coefficients signatures.}
	\label{figure_17}
\end{figure}

Thus, regarding the dominant spiking channel, a behavior signature is generated in the local channel data and the affected adjacent channel simultaneously as described in the previous work,~\citet{zaghloul2015adaptive}. The block diagram of the Signal Conditioning Stage is demonstrated in Fig~\ref{figure_17}.

Each generated signature of the input sliding window slot is represented as a 165-bit word that contains the major characteristics of the signal coming for the DWT coefficients. They consist of the local signature (8$\times$16 bit) and global signature (4$\times$8 bit) plus another five control bits divided into three strings that are used for: priority, ordering and control bits, respectively. Therefore, in order to achieve a real-time matching evaluation of the incoming data behavior, the generated signature is then fed directly into the next stage of the Adaptive Signal Analysis.

\subsubsection{Adaptive Signal Analysis}

In this stage, the signatures and the input signals are parallel fed in into two main units: the Seizure Detection and the Seizure Early Prediction units. The seizure detection unit we used is an enhanced version (tolerant of artifact noise) of the technique described in~\citet{ponten2010feasibility}. 

\begin{figure*}%*[t]
	\centering
	\includegraphics[width=11cm,height=9cm]{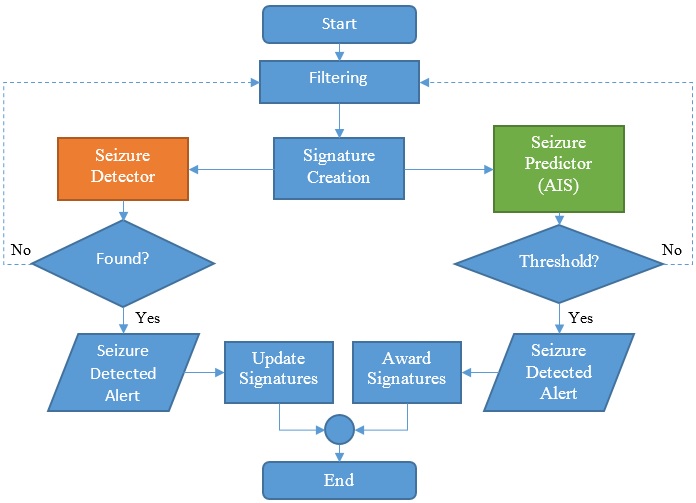}
	\caption{The proposed system operation algorithm flowchart.}
	\label{figure_18}
\end{figure*}

The system detects the diversity for the standard deviation with combining the resulted value matrix of size NxN, where N is the number of channels, then calculates the max of the signal likelihood values. Then the likelihood value is compared to the adaptive threshold and duration (number of clock cycles of occurrence) to detect a seizure.  The Detection Unit can also be done via the AIS algorithm which could enhance the detection accuracy by 17\%; however, this will cause an increase in the chip area by 57\% and its role as a backup plan strategy will not be used with respect to the Early Predicting Unit. The Prediction Unit principal engine of the parameterized AIS algorithm is shown in Fig~\ref{figure_18}.

Due to the fast processing of the comparator and the low power ALU, described in~\citet{al2012green}, the speed was compromised to achieve a smaller area. The early prediction function unit can be summarized as follows:

\begin{itemize}
	\item \textbf{AIS Detection}: matches the incoming signatures with signatures from the SLT, based on MSE value inspired by the negative selection Algorithm. In case of detection, the winning signature will be placed on the top of the SLT stack. However, in the case of a new signature found, it will be appended to the SLT.
	\item \textbf{Population Manager}: the core of the AIS that assigns the priority, the sensitivity parameters, and triggers the Signal Mutation process. It is the signature population manager that guarantees the diversity between the signatures by performing the mutation and changing the priorities.
	\item	\textbf{Signature Mutation}: fires in two cases: the sustained detection to the top of the stack, or the sliding window count W was reached. In both cases, the mutation happens to the lookup table in order to allow the Population Manager to process the new signatures and adapt new priority values.
\end{itemize}

\begin{figure*}%*[t]
	\centering
	\includegraphics[width=11cm,height=7cm]{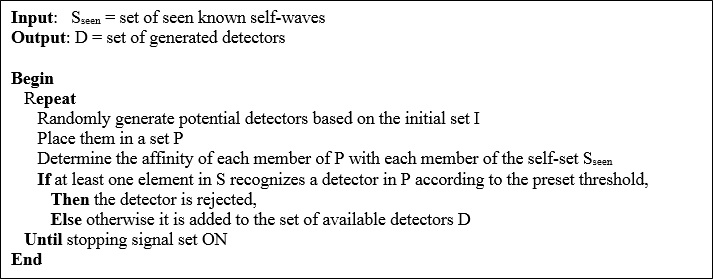}
	\caption{The pseudo code of the proposed algorithms.}
	\label{figure_19}
\end{figure*}

The main detection engine algorithm is based on the negative selection algorithm idea where it tries to find the abnormality in the system for every rouge input (antigen). The pseudo code for the algorithms shown in Fig~\ref{figure_19}. The starting point of this algorithm is to produce a set of self-waves, S, that define the normal state of the system. The task then is to generate a set of detectors, D, that only bind/recognize the complement of S.

\begin{figure*}%*[t]
	\centering
	\includegraphics[width=11cm,height=7cm]{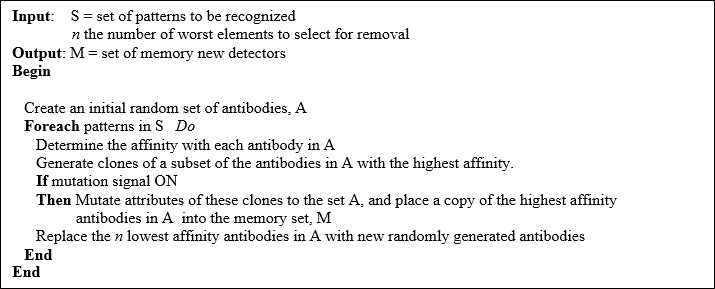}
	\caption{The pseudo code for the Clonal selection algorithm.}
	\label{figure_20}
\end{figure*}

The algorithm’s clonal selection theory has been used as inspiration for the development of AIS that perform computational optimization and pattern recognition tasks. In particular, inspiration has been taken from the antigen-driven affinity maturation process of B-cells, with their associated mutation mechanism. These AIS also often use the idea of memory cells to retain good solutions to the problem being solved. The pseudo code for the clonal selection is shown in Fig~\ref{figure_20}.

The affinity $d$ measurement is determined using the Euclidean distance of the antigen with the detectors that are defined by:
\begin{equation}
d(S,D) = \sum_{i=1}^{n} (S_i -D_i)^2
\end{equation}

Using the adaptive behavior of the architecture with the mutation shuffling dramatically reduces the local minima scenario and increases the true negative detection (e.g. specificity) from the usual pattern recognition scenario by more than 12\%. 

\subsubsection{Decision Making Unit}

This is responsible for sending the warning signal of the predicted seizure. It is wanted for existing seizures or transferring the EEG data to the external monitoring device or warning appliance. It compares the output of the seizure detection unit and the seizure prediction unit with respect to the time and the current sliding window, and it chooses to either warn for the upcoming possible pre-ictal, or alarm for the existing seizure-ictal. Also, this unit reads the threshold, which is the signal sensitivity parameters from that external device.

\subsubsection{The Overall Design}

\begin{figure*}%*[t]
	\centering
	\includegraphics[width=11cm,height=10cm]{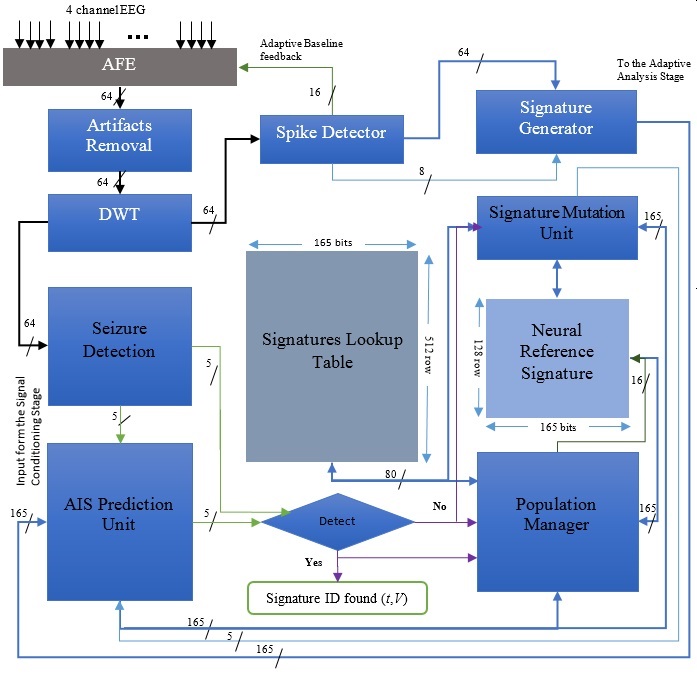}
	\caption{The proposed system block diagram.}
	\label{figure_21}
\end{figure*}

Finally, by combing the three phases of the operation, we get the overall system block diagram that is shown in Fig~\ref{figure_21}, including the EEG Input registers and the AFE units. Then, as soon a seizure prediction hits the threshold, a warning signal is generated and sent to the output unit stating the time and the window of the EEG that caused the firing to the external device. On the other hand, if a seizure onset is detected, an alarm is sent to the external device and an update to the population manager is sent to either add the previous signature if it does not exist or increase the priority weight of the existing signature.

We implemented the algorithm in C/C++ and Matlab to simulate the operation, then we designed the system in Simulink for Verilog for simulation and future hardware implementation. The simulation and results are presented in the next subsection. We implemented the algorithm in C/C++ and Matlab to simulate the operation, then we designed the system in Simulink for Verilog for simulation and future hardware implementation. The simulation and results are presented in the next subsection.

\subsection{Simulation Results}

\begin{figure}%*[t]
	\centering
	\includegraphics[width=8.5cm,height=6cm]{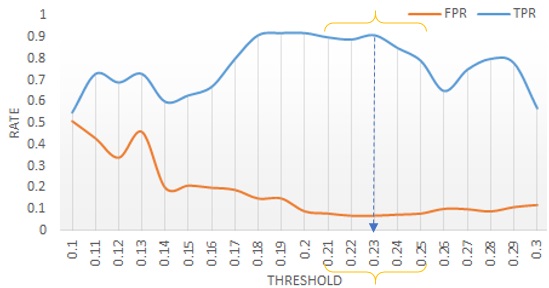}
	\caption{The correlation between the threshold, FPR, and TPR.}
	\label{figure_22}
\end{figure}

Initially, the algorithm was experiencing a training phase. We used Matlab, Simulink, EEGLAB, and WaveClus for the initial simulation of the algorithm and decided on the initial values of the threshold window width and the parameters wire data width. To select the width and the constant parameters of the thresholds, we compare them to the FPR. After 5 min of simulation and by studying the different threshold values of the accuracy of the system, we selected the initial value of the combined likelihood value between 0.20 and 0.25 with a mean of 0.23. This is slightly higher than the previous work by 33.3\% due to deployment of the AR unit; the value was selected to optimize the FPR, as shown in Fig~\ref{figure_22}, where we can find that the relatively stable lower FPR and the highest TPR in the approximate range, as described in~\citet{ponten2010feasibility} and~\citet{haas2007strategies}, where the initial threshold was selected to be. We are aware it is not the optimal mathematical solution for the minimization of the FPR and the maximization of the TPR to be optimized, but this approximate solution provides a decent initial value for the threshold, where the rest is up to the doctor for fine tuning of the system regarding each individual patient or special situation.

\begin{figure}%*[t]
	\centering
	\includegraphics[width=8.5cm,height=6cm]{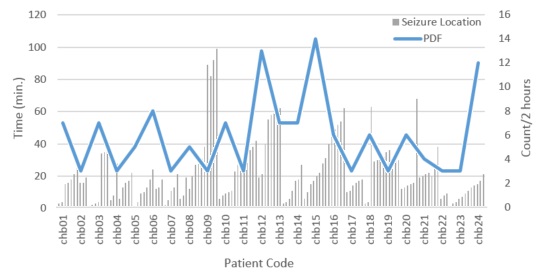}
	\caption{Seizure data of the density of seizure occurrence during two hours recording.}
	\label{figure_23}
\end{figure}

The next step was to generate the initial population of signatures of the AIS. We used the data from CHB-MIT lab, AL Goldberger et al., 2000, where an experiment was done on 24 patients at a Boston Hospital in December 2010 for an average of two hours. The data and statistics graph from this study is shown in Fig~\ref{figure_23}. We then randomly selected 8 patients’ data to act as training data. Next, we used the rest of the data for the testing and verification phase containing seizure and seizure free recordings. The data consist of scalp EEG recordings of pediatric subjects with intractable seizures. Subjects were monitored for up to several days following the withdrawal of anti-seizure medication in order to characterize their seizures and assess their candidacy for surgical intervention; the patient’s statistics are shown Table~\ref{table_1} and are according to the expert’s visual inspection of the EEG recordings.

\begin{table}[htbp]
	\caption{The characteristics of the selected patients for learning phase}
	\begin{center}
		\begin{tabular}{lllllllll}
			\hline
			\textbf{Patient Code}&\textbf{P1}&\textbf{P2}&\textbf{P3}&\textbf{P4}&\textbf{P5}&\textbf{P6}&\textbf{P7}&\textbf{P8}\\
			\hline
			\textbf{Gender}&F&M&F&F&M&F&M&F\\
			\hline
			\textbf{Age (years)}&11&11&14&2&16&18&3&14.5\\
			\hline
			\textbf{E.S./2 hours}&7&3&7&13&14&6&7&3\\
			\hline
		\end{tabular}
		\label{table_1}
	\end{center}
\end{table}

These parameters were used to determine the initial value for the configuration parameters that define the number of the initial antibody set, the number of the generated clones of each antibody, the gene count (i.e., matching samples), the mutation cycle length, the value of the threshold to the antigen and the signature selection, shown in Table~\ref{table_2}.

\begin{table}[htbp]
	\caption{Selected patients for learning phase characteristics}
	\small
	\begin{center}
		\begin{tabular}{ll}
			\hline
			\textbf{Parameter}&\textbf{Value}\\
			\hline
			Number of antibodies&50\\
			%	\hline
			Number of clones &25\\
			%	\hline
			Genes&255\\
			%	\hline
			Mutation cycles&83\\
			Remove threshold&0.3\\
			Selection threshold& 0.01\\
			Diversity& 0.64\\
			\hline
		\end{tabular}
		\label{table_2}
	\end{center}
\end{table}

\begin{figure}%*[t]
	\centering
	\includegraphics[width=8.5cm,height=6cm]{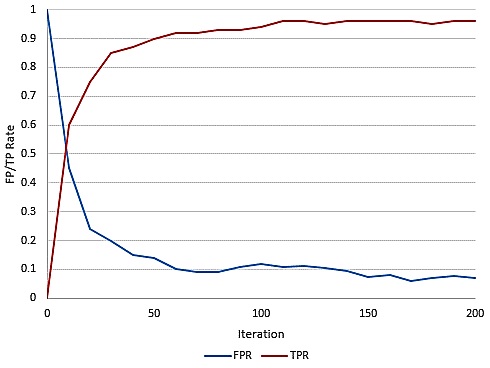}
	\caption{The learning curve of detector initial threshold (td) of FPR and TPR.}
	\label{figure_24}
\end{figure}

The learning curve of the initial likelihood threshold of the detection algorithm with respect to the FPR and the TPR was set using two hundred input segments, each of 5 sec duration that included a seizure onset event for not less than 3sec each. This was done to ensure a variable location of the ictal event within the sample space for the detection mechanism input. The results are shown in Fig~\ref{figure_24}.

\begin{figure}%*[t]
	\centering
	\includegraphics[width=8.5cm,height=6cm]{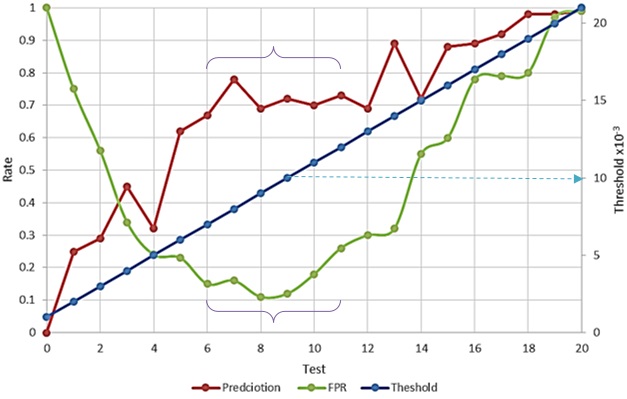}
	\caption{The initial prediction threshold.}
	\label{figure_25}
\end{figure}

For the prediction unit, the threshold ($tp$) of the detector’s likelihood of prediction was set empirically with a twenty test segment, which was taken from the previous data sample for 20 sec right before the seizure ictal event. The event was marked via the visual inspection of the neuroscientist (given with the data). The initial threshold was set via the imperial test, which was similar to the method used to determine the initial value of $td$ in regards to the correlation between the FPR and the TPR, as shown in Fig~\ref{figure_25}.

The values of the $td$ and $tp$ are initially set to 0.23, and 0.09 respectively. These values were chosen via the imperial test and serve as a guideline for the system to function initially. However, the endpoint management application will give the ability for the doctors of the ICU to tune these parameters to the sensitivity of the warning and the alarm thresholds which suits their need for every different case, situation or patients’ needs.
After the training phase was completed, we used several patients’ EEG data, excluding the previously selected eight patients who were used for training. The simulation ran for a period of 5 min each, where the prediction time threshold was initially set from 10-15 sec, and the system was able to predict and detect seizures for most of the selected patients with a sensitivity of 71\% and an accuracy of 91\%. The results of the experiments are shown in Table~\ref{table_3}.

\begin{table*}[htbp]
	\caption{Selected patients for learning phase characteristics}
	\small
	\begin{center}
		\begin{tabular}{lllllllll}
			\hline
			\textbf{Patient Code}&\textbf{P1}&\textbf{P2}&\textbf{P3}&\textbf{P4}&\textbf{P5}&\textbf{P6}&\textbf{P7}&\textbf{P8}\\
			\hline
			\textbf{Gender}&M&F&F&M&F&F&F&F\\
			\hline
			\textbf{Age (years)}&22&17&10&7&3.5&12&9&6\\
			\hline
			\textbf{E.S./2 hours}&3&5&3&6&5&3&3&12\\
			\textbf{Predicted}&5&6&3&8&9&4&4&14\\
			\hline
			\textbf{Detected}&4&5&3&6&10&4&3&11\\
			\hline
			\textbf{T-to-S (sec)}&12.4&16.32&16.6&10.8&13.4&18.3&17.2&11.6\\
			\hline
		\end{tabular}
		\label{table_3}
	\end{center}
\end{table*}

After the simulation step, a verification procedure took place. We have compared the proposed early prediction algorithm to a simple random guess from a normal distribution. The test that was proposed and used to verify the prediction step was performed on logical decisions bases, not a random guess. After that, we also compared our proposed System II to the previously proposed architecture Arch1, and Arch2, proposed in~\citet{rajdev2010real} and~\citet{ponten2010feasibility}, respectively.

\begin{figure}%*[t]
	\centering
	\includegraphics[width=8.5cm,height=6cm]{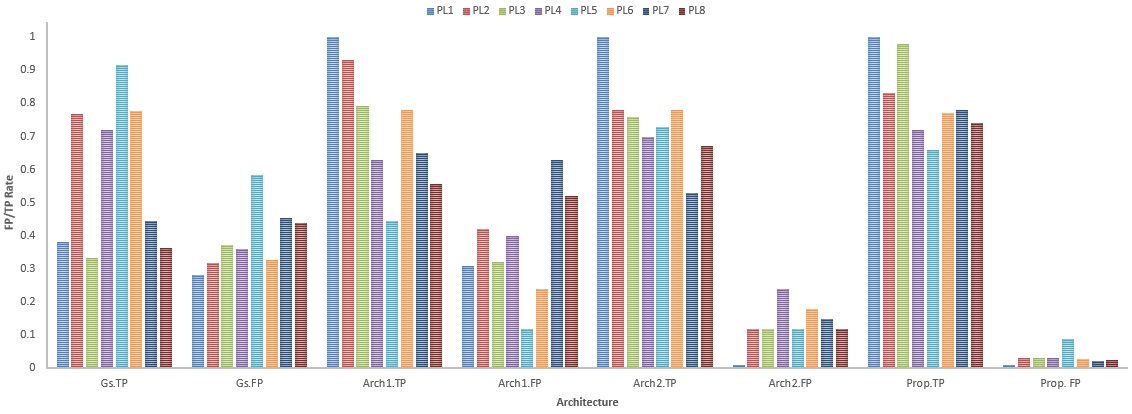}
	\caption{The comparison of FP and TP for ``System2'', previous work and random guess.}
	\label{figure_26}
\end{figure}

Further, our proposed algorithm outperforms the area above the curve of the random guess, and compared to the previous work, we achieved a lower false positive alarm by 36\%. We also produced an average earlier prediction of 14.57 sec, which is more than twice as fast as the state-of-the-art performance, shown in Fig~\ref{figure_26}.

From the previous graph, shown in Fig~\ref{figure_26}, we can see that our results were verified to be producing better FPR and TPR than a random guess, existing TPR and lowering FPR than the state-of-the-art architectures. The most improvement was in the FPR; that is the most important aspect of the early prediction domain.

\subsection{Results Analysis}

In this subsection, we are discussing and commenting on the result that we got for the simulation and the comparison with the previous work and the state-of-the-art architectures. First, all of the previous work and the proposed work exceed the performance of the random guess which makes them all a selective candidate for future implementation, although, Arch. 1 and Arch. 2 provided an acceptable average result (following the prediction standards stated by~\citet{jerger2001early}). 

We have noticed that Arch.1 has the lowest average performance compared to Arch.2 and our proposed method. After analyzing the situation, it turns out that Arch.1 would have performed better it the authors would have proposed a filter in the system as it is in the case in the other two systems, which makes the results biased for that architecture. However, we compared our work with the stated systems exactly as is.
Another observation was done due to the lower performance that we get with PT3, as there were more false positive predictions and false detections than the mean values. Therefore, we studied the EEG and the data for that specific case in more details, and it turns out that for this specific case there were different unnatural artifacts that affect the detection and the prediction system due to the patient’s condition and age. This was the reason for the abnormal behavior of the system.

On the other hand,~\citet{di2010seizure} claimed to describe that a dog may warn of impending psychogenic non-epileptic seizures (PNES), saying that such a dog’s companionship may cut down on seizure frequency. Also, they mentioned that the ability of a dog to obtain help during or after a seizure could prove lifesaving. Further, they showed in their study that a patient was “alerted by his dog 7 minutes prior to having psychogenic seizures.” However, the experiment was carried with a success rate of four out of six parliaments. Additionally, a strong criticism by~\citet{doherty2007wag} addresses that study. Thus, a brief, comprehensive summary of the state-of-the-art systems, existing claimed methods, and the nerve stimulation implant solution is listed in addition to our proposed system in Table~\ref{table_4}.

\begin{table*}[htbp]
	\caption{A comprehensive summary of the prediction systems}
	%	\small
	\begin{center}
		\begin{tabular}{llllll}
			\hline
			\textbf{Comparison}&\textbf{Claim}&\textbf{SNA}&\textbf{Arch1}&\textbf{Arch2}&\textbf{Proposed}\\
			\hline
			\textbf{Detection}&N/A&Yes&Yes&Yes&Yes\\
			\hline
			\textbf{Prediction}&Yes&No&No&Yes&Yes\\
			\hline
			\textbf{Input}&N/A&Implant&Implant&Scalp EEG&Scalp EEG\\
			\hline
			\textbf{Average time}&7 min&--&--&6.4 sec&\textbf{}14.57 sec\\
			\hline
			\textbf{Detected}&N/A&65\%&65\%& 91\% & 92\%\\
			\hline
			\textbf{Prediction Accuracy}&66.7\%&--&--&63\%&72\%\\
			\hline
			\textbf{Hardware}&--&Implant&Yes&Yes&Yes\\
			\hline
			\textbf{Learning}&N/A&N/A&Threshold Adaptation& Static SVM Seizure patterns& Adaptive AIS with adaption\\
			\hline
			\textbf{Filtering}&N/A&N/A&No&Frequency based for High Freq. Artifacts& Cauchy Based (EOC, EMG)\\
			\hline
		\end{tabular}
		\label{table_4}
	\end{center}
\end{table*}

As a summary, we have shown, illustrated and verified that our novel proposed system is capable of early prediction of several types of epileptic seizure, with a prediction period average of 14.57 sec, twice that of the state-of-the-art devices, with an average accuracy of 72\% and a lower FPR than the current system by 75\%. However, there are several disadvantages or undeveloped situations which the proposed system at this stage cannot handle. The first issue is the abnormal artifacts that usually exit with several infants who suffer from a seizure. Infant Spasm Syndrome, and Febrile Seizure, which usually occur in children aged 3 months to 5 years, are examples. They usually occur with a high fever. However, such cases are considered to be rare (i.e. 2\% to 5\% of all the children).
The second issue is that there is a relatively short prediction time compared to the claimed dogs’ ability to predict seizures (if this claim could be proven). However, we believe that the advances in neurosciences can bring us more clues and more input data signal markers that we can extract to predict a better pattern of information for a more accurate prediction within the EEG data.

\section{Conclusions and Future Works}

We propose an early prediction and detection system for epileptic seizures. Regarding the problem statement for the epilepsy detection issue, the proposed system was able to solve almost all of the pending issues with the detection technology. It proposed a novel – not only a seizure detection system – but early epilepsy prediction feature that uses a parameterized AIS based adaptive prediction system that was verified and outstood the state-of-the-art architectures through prediction accuracy by 114\%, the false positive rate by (75\%) and the average prediction time by more than twice the current prediction time.

The main detection problem and issue that were discussed in the introduction to the surgical procedures, including the power and battery replacement issue, the noise proof design for detectors and predictors issue, and the high false alarm rate issues, where solved via the proposed wearable and disposable wireless band with minimally invasive dry EEG sensors for signal acquisition system architecture. The system is EOC and EMG artifacts-proof due to the support of the Cauchy based artifact filter, and the accurate parameterized AIS prediction mechanize, respectively. Making the proposed system not only a promising early prediction system architecture but a novel horizon for BCI based solution for many brain disorders’ early prediction, diagnoses, and detection systems.

Artifact noise reduction modeling and filtering, especially the non-stationary noise, is a very important and challenging phase. Together with the deployment of wireless EEG data, acquisition sensors can bring a new era of non-invasive BCI implants. Additionally, if it was possible to find a fine brain wave detector, we could get a new set of BCI techniques and methods that do not require an implanted chip and can read the brain’s neural activity completely remotely.

Another idea is to extend the AIS prediction algorithm in early diagnostics mechanisms that will be based on more clues of the scalp EEG data. More optimized and specialized recognition markers not limited to prediction but could perform an early diagnosis of several mental disorders; for example, early prediction of Alzheimer Disease, which could enhance the diagnosis and prediction control modeling. This could help to open new horizons for BCI implementation of the cypher medical domain for better prediction, detection and even prevention application for a better tomorrow for the whole of humanity.

\section{Acknowledgment}

The authors would like to express special thanks to Philip, Inota, and Raj for their help with the Matlab simulation and implementation; great appreciation for Jaqueline Hebert for the review.

	%%%%%%%%%%%%%%%%%%%%%%%%%%%%%%%%%%%%%%%%%%%%%%%%%%%%%%%%%%%%%
	\bibliographystyle{plainnat}%{rusnat}%{rusnat}%{plainnat}
	\bibliography{eeg_bci}
	
%	\EOD

\end{document}